# Suppression of 1/$f$ Noise in Near-Ballistic $h$-BN-Graphene-$h$-BN Heterostructure Field-Effect Transistors

Maxim A. Stolyarov[1], Guanxiong Liu[1], Sergey L. Rumyantsev[2,3], Michael Shur[2] and Alexander A. Balandin[1, *]

[1] Nano-Device Laboratory (NDL) and Phonon Optimized Engineered Materials (POEM) Center, Department of Electrical and Computer Engineering, Bourns College of Engineering, University of California – Riverside, Riverside, California 92521 USA

[2] Center for Integrated Electronics and Department of Electrical, Computer and Systems Engineering, Rensselaer Polytechnic Institute, Troy, New York 12180 USA

[3] Ioffe Physical-Technical Institute, Russian Academy of Sciences, St. Petersburg, 194021 Russia

## Abstract

We have investigated low-frequency 1/$f$ noise in the boron nitride – graphene – boron nitride heterostructure field-effect transistors on Si/SiO$_2$ substrates ($f$ is a frequency). The device channel was implemented with a single layer graphene encased between two layers of hexagonal boron nitride. The transistors had the charge carrier mobility in the range from ~30000 to ~36000 cm$^2$/Vs at room temperature. It was established that the noise spectral density normalized to the channel area in such devices can be suppressed to ~5×10$^{-9}$ μm$^2$ Hz$^{-1}$, which is a factor of ×5 – ×10 lower than that in non-encapsulated graphene devices on Si/SiO$_2$. The physical mechanism of noise suppression was attributed to screening of the charge carriers in the channel from traps in SiO$_2$ gate dielectric and surface defects. The obtained results are important for the electronic and optoelectronic applications of graphene.

The most realistic of the proposed electronic applications of graphene are those that do not seriously suffer from the absence of an energy bandgap but rely mostly on graphene's high electron mobility, thermal conductivity, saturation velocity, and a possibility of tuning the charge carrier concentration over a wide range [1, 2]. These applications are in analog electronics [3-5], high-frequency graphene devices for communications [6, 7], and terahertz plasmonic devices [8, 9] which rely on its excellent electron mobility and saturation velocity, as well as in chemical and biological sensing enabled by the ultimately high surface-to-volume ratio and the precise control of the carrier concentration [10-13]. For all these applications, the low-frequency electronic $1/f$ noise is a crucial performance metric (here $f$ is the frequency). The low-frequency noise, usually found at frequencies below 100 kHz, limits the sensitivity and selectivity of all the sensors that rely on an electrical response. It is also responsible for the dominant contribution to the phase noise of the communication systems even when they operate at much higher carrier frequency [14-16]. From the fundamental physics point of view, graphene, as a truly two-dimensional material system, presents an interesting testing ground for theories describing the origin and mechanisms of $1/f$ noise [17-19].

Owing to the importance of the subject, there have been numerous reports on $1/f$ noise in graphene devices [17-30]. Despite some data scatter due to unavoidable differences in graphene and device quality, most of the studies agree on the following characteristics of low-frequency noise in graphene. The low-frequency noise spectral density, $S_I$, in graphene devices is proportional to $I^2$ (here $I$ is the drain – source current). The latter implies that the electrical current does not drive the fluctuations but merely makes them visible as in other homogeneous conductors [31]. Although both the graphene layer itself and metal contacts contribute to the $1/f$ noise, the dominant contribution mostly comes from the graphene channel itself. The results obtained by different groups for micrometer size graphene devices on Si/SiO$_2$ substrates put the current normalized spectral density $S_I/I^2$ in the range $10^{-9}$ to $10^{-7}$ Hz$^{-1}$ at $f = 10$ Hz [17]. A more informative characteristic of $1/f$ noise level in two-dimensional (2D) materials is the noise spectral density normalized to the device channel area, which we denote as parameter $\beta=(S_I/I^2)(W\times L)$, where $W$ is the channel width and $L$ is the channel length. Independent studies established that $\beta$ parameter is in the range from ~ $10^{-8}$ to $10^{-7}$ $\mu m^2$ $Hz^{-1}$ for micrometer scale graphene devices on Si/SiO$_2$ substrate. Another important finding reported by several groups [17, 19, 28-30] was that 1/$f$ noise in graphene does not follow the McWhorter model [32] conventionally used for description of noise in Si complementary metal-oxide-semiconductor (CMOS) transistors and field-effect transistors (FETs) made of other conventional semiconductors [15].

Different mechanism of noise in graphene calls for investigation of the noise reduction techniques that can be effective for the specific material system. In this Letter, we report on the low-frequency noise in the hexagonal boron nitride – graphene – boron nitride (*h*-BN-G-*h*-BN) heterostructure field-effect transistors (HFETs) on Si/SiO$_2$ substrates. The mobility in back-gated and top-gated graphene devices on Si/SiO$_2$ substrates used for noise studies previously was in the range from 1500 to 7000 cm$^2$/Vs at room temperature (RT). In our HFETs graphene channel is screened from defects by the hexagonal BN cap and barrier layers. The latter resulted in RT mobility in the back-gated graphene HFETs in the range from ~30000 to 36000 cm$^2$/Vs and allowed us to study the low-frequency noise in the supported graphene devices operating in the near-ballistic transport regime [33].

The specific structure of *h*-BN-G-*h*-BN heterogeneous device channel was selected following the reports of mobility enhancement in graphene devices on *h*-BN substrate [34-36]. We modified the device design by using a thicker *h*-BN barrier and cap layers for better screening from the defects. The devices for this study were fabricated in the following steps. First, *h*-BN layers (thickness $H$~30 nm) were mechanically exfoliated on top of p-doped Si/SiO$_2$ wafer (300 nm of SiO$_2$). Graphene layers were prepared by the same procedure on another Si/SiO$_2$ substrate. Thin viscoelastic materials (Gelpak) adhered to glass slides were used as transparent stamps for the layer transfer. The stamps were spin-coated (Headway SCE) with polypropylene carbonate (PPC). Second, the stamp with PPC was brought into contact with the *h*-BN layer on a substrate using a micromanipulator. The stage was heated to 40°C allowing for adherence of *h*-BN crystal with subsequent lifting of the stamp with attached *h*-BN layer. Third, the micromanipulator was used for careful positioning of *h*-BN layer over the graphene layer. Repeating these steps, *h*-BN – graphene – *h*-BN stacks were formed creating the desired heterostructure. The completed heterostructure was released by heating the stage to an appropriate temperature onto the degenerately doped p-type Si/SiO$_2$ substrate. Finally, the PPC layer was washed out with acetone to leave the assembled stack on the substrate. No cleaning treatments like thermal annealing have been used before and during the assembly process as well as in the post-fabrication stage. Raman spectroscopy (Renishaw InVia) was used to determine the number of atomic planes in the exfoliated graphene samples and to verify the quality of the selected graphene and *h*-BN layers. Figure 1 (a-b) shows an optical microscopy image of graphene on top of *h*-BN layer and Raman spectrum of the exfoliated graphene. The ratio of the intensity of G peak to the intensity of 2D band and deconvolution of the 2D band prove that the device channel is made of single layer graphene.

[Figure 1 (a-b): optical image and Raman]

The fabricated heterostructures were spin coated and heated with a positive resist polymethyl methacrylate (PMMA) two times. Patterning of the assembled stacks was accomplished by electron beam lithography (EBL) (LEO Supra). In order to expose encapsulated graphene edges the assembled stacks were selectively etched with sulfur hexafluoride (SF6) gas on an inductively coupled plasma system (Oxford Plasmalab) into conventional Hall bar geometries. The samples were rinsed with acetone to remove the resist mask. After the repeated PMMA spin coating procedures, the electrical contacts were patterned with EBL. Immediately before metallization, the graphene edges were exposed to O$_2$ plasma to improve bonding and increase transmission [37, 38]. The metal leads (10 nm Cr / 100 nm Au) were deposited by electron beam evaporation (Temescal BJD). The electrical contacts were made to Cr adhesion layers because the Cr work function is ~0.16 eV lower than that of graphene [39]. The fabricated three-dimensional (3D) Cr/Au electrodes touched the 2D graphene monolayer along the one-dimensional (1D) graphene edge in these devices. This "1D contact" approach is typically advantageous in comparison to conventional "2D contacts" in the sense of separating the layer assembly and metallization processes, lower contact resistance [38]. The schematics and optical microscopy image of a representative device are presented in Figure 2. Total of eight devices were studied in this work.

[Figure 2: device schematic and microscopy image]

Figure 3 shows the current-voltage (I-V) characteristics of a representative *h*-BN-G-*h*-BN HFET with *L*=9.45 μm. The effective mobility, $\mu_{eff}$, was determined from the channel resistance using the expression

$$\mu_{EFF} = \frac{L_G}{R_{EFF} C_G (V_{GS} - V_D) W} \, , \tag{1}$$

where $L_G$ is the gate length, $C_G$ is the gate capacitance per unit area, $R_{EFF} = \frac{R_{DS} - R_C}{1 - \sigma_0 (R_{DS} - R_C)}$, $\sigma_0$ is the conductivity at the voltage corresponding to the charge neutrality point, $R_C$ is the sum of the drain and source contact resistances, and $R_{DS}$ is the measured drain-source contact resistance. All our measurements were performed in the linear regime at very small currents so that the external $V_{GS}$ was approximately equal to the intrinsic source-gate voltage. The field-effect mobility, $\mu_{FE}$, was determined from the transconductance, $g_{m0}$, in the linear regime using the expression

$$\mu_{FE} = \frac{g_{m0}}{C_G (V_{DS} - I R_C)} \frac{L_G}{W} \, , \tag{2}$$

where $V_{DS}$ is the drain-source voltage. In the linear regime at small drain voltages the internal transconductance was found from

$$g_{m0} \approx g_m \left(1 + \frac{R_C}{R_{EFF}} + R_C \sigma_0 \right), \tag{3}$$

where $g_m$ is the external transconductance. Both the effective and field-effect mobility extractions gave consistent results and the charge carrier mobility was determined to be greater than 30000 cm$^2$/Vs at RT and for the carrier concentration of $7 \times 10^{11}$ cm$^{-2}$. The low-temperature (*T*=77 K) mobility values reached 100,000 cm$^2$/Vs.

An estimate for the contact resistance, $R_C$, was obtained by plotting the drain-to-source resistance, $R_{DS}$, versus $1/(V_G - V_D)$, where $V_G$ is the gate bias and $V_D$ is the Dirac (charge-neutrality point) voltage. An extrapolation of this dependence to $1/(V_G - V_D) = 0$ provides the sum of the drain and source contact resistance $R_C$. For the studied devices, the value of the contact resistance per unit width, $R_{c0} = R_C \times W/2$ was ~ 277 Ωμm. To estimate the charge carrier mean free path (MFP), $\Lambda$, we used a conventional relation between the mobility, $\mu$, and electrical conductivity $\sigma = en\mu = (2e^2/h)k_F\Lambda$, where $k_F = (\pi n)^{1/2}$ is the Fermi wave vector in 2D graphene, $h$ is the Plank's constant, $e$ is the charge of an electron and $n$ is the carrier concentration. For $n=2\times10^{12}$ cm$^{-2}$, from the expression $\Lambda = (h/2e)\mu(n/\pi)^{1/2}$ we obtained $\Lambda\approx0.311$ μm. For devices with $W\approx1$ μm and $L$ in the range from 2.5 to 9.45 μm the electron transport is not yet ballistic, but it approaches this regime. As predicted in [40], the unique features of the near ballistic response could be revealed by studying "ringing" of the transistor response to short terahertz pulses.

[Figure 3: I-V characteristics]

The low-frequency noise was measured using an in-house built experimental setup with a spectrum analyzer (SRS FFT). The devices were biased with a "quiet" battery - potentiometer circuit. Details of our noise measurement procedures have been reported by some of us elsewhere [18, 19, 22]. Figure 4 shows representative normalized noise spectrum density, $S_I/I^2$, for one of the tested devices. The noise is of true $1/f^\gamma$ type with $\gamma$ varying from 0.95 to 1.2 with an average $\gamma = 1.09$ for a device with the channel $W\times L=1.16\times9.45$ μm$^2$. For the device with the channel $W\times L=1\times3.21$ μm$^2$, the extracted $\gamma$ was in the range from 0.84 to 1.27 with an average value of $\gamma=1.02$. Table I lists the $\gamma$ values for two representative devices. No trend in $\gamma$ dependence with the device channel area or gate voltage, which would suggest non-uniformity of the charge trap distribution [15], was observed. The noise spectra of all examined devices revealed no traces of the generation-recombination (G-R) noise. The noise spectrum density of the high-mobility $h$-BN-G-$h$-BN HFETs revealed strongly non-monotonic gate-bias dependence, which is contrast to that described by the McWhorter model in Si CMOS devices [32].

[Figure 4: Noise vs frequency]

**Table I: Parameter $\gamma$ for low-frequency noise in BN-Graphene-BN HFETs**

| $V_G$ (V) | -60 | -30 | -10 | -8 | -5 | 0 | 5 | 10 | 30 | 60 |
|---|---|---|---|---|---|---|---|---|---|---|
| $\gamma$ (device A) | 1.16 | 1.04 | 0.96 | 1.19 | 0.95 | 1.16 | 0.97 | 1.12 | 1.12 | 1.20 |
| $\gamma$ (device B) | 1.27 | 0.84 | 0.95 | 1.04 | 1.02 | 1.02 | 0.96 | 0.94 | 1.04 | 1.05 |

To better elucidate the non-monotonic type of the noise gate bias dependence we calculated the noise amplitude as $A = \frac{1}{Z}\sum_{i=1}^{Z} f_i S_{Ii}/I^2$, which is a noise characteristic analogous to the normalized noise spectral density $S_I/I^2$ but averaged over several frequencies (here $Z$ is the number of the frequency data points). Figure 5 shows the noise amplitude in our $h$-BN-G-$h$-BN HFET as a function of $V_{GS}$-$V_D$ ($V_D$ is the Dirac voltage). For comparison, we also show noise amplitude in conventional non-encased graphene FET on Si/SiO$_2$ reported in Ref. [28]. The non-encased graphene FET had mobility less than 3000 cm$^2$/Vs. In our high-mobility $h$-BN-G-$h$-BN HFETs the minimum of the noise amplitude was achieved near the Dirac point, similar to that in the conventional graphene FETs [17, 28].

[Figure 5: Noise amplitude]

The parameter $\beta = (S_I/I^2)(W \times L)$ is a better characteristic of 1/$f$ noise in 2D materials than Hooge parameter introduced specifically for volume noise [22]. For conventional graphene devices on Si/SiO$_2$ substrates $\beta$ parameter is ~10$^{-8}$ to 10$^{-7}$ μm$^2$ Hz$^{-1}$ for micrometer-scale channels [17]. In our high-mobility $h$-BN-G-$h$-BN HFETs $\beta$ was determined to be in the range from 5×10$^{-9}$ to 2×10$^{-8}$ μm$^2$ Hz$^{-1}$. At small gate biases the noise level was typically below 10$^{-8}$ μm$^2$ Hz$^{-1}$ in $h$-BN-G-$h$-BN HFETs. On average, 1/$f$ noise in our devices was suppressed by a factor of ×5 – ×10 as compared to that in non-encapsulated graphene devices on Si/SiO$_2$. This is a substantial reduction, which can have practical implications.

We now turn to explanation of a potential mechanism of noise reduction in encased graphene channel devices. It is generally accepted now that the low frequency 1/*f* noise can be either due to the number of charge carrier fluctuations or due to their mobility fluctuations or both. Depending on which term dominates one distinguishes the mobility fluctuation or the carrier number fluctuation mechanism of 1/*f* noise [15]. In Si and other metal-oxide-semiconductor field-effect transistors (MOSFETs) the carrier number fluctuations usually dominate and such type of the noise is well described by the McWhorter model [32]. The studies that investigate noise in graphene under irradiation [19] and magnetic field [41] suggested that the mechanism of 1/*f* noise in graphene is more similar to the mobility fluctuations mechanism (like that in metals).

Owing to graphene's atomic thickness and the fact the mobility is limited by scattering from defects and impurities in $SiO_2$ [42-46], the mobility fluctuations will be due to the fluctuations in the scattering cross-sections of defect states in $SiO_2$ gate dielectric. For this reason, irrespective of the specific noise mechanism – carrier number or mobility fluctuations – screening electrons in graphene channel from the defect states in $SiO_2$ by introducing *h*-BN barrier layer with the thickness of 30 nm should reduce the noise. This conclusion is consistent with reports that 1/*f* noise was reduced in the suspended graphene devices [23]. It is interesting to note that suspended graphene device reported in Ref. [23] had $\beta \sim 6 \times 10^{-9}$ µm$^2$/Hz which is approximately the same noise level as in our encased graphene channel HFETs. While most of noise reduction is likely related to screening of the graphene channel from traps in $SiO_2$ it is possible that capping graphene with *h*-BN also helps to reduce the noise. It has been shown that the environmental exposure and device ageing increase the level of 1/*f* noise in graphene devices [17, 22]. Organic residue and other contaminants on the surface can create either trapping centers for electrons in the channel (carrier number fluctuation noise) or scattering centers (mobility fluctuation noise).

In conclusion, we investigated the low-frequency 1/*f* noise in the *h*-BN-graphene-*h*-BN HFETs on Si/$SiO_2$ substrates. The heterostructure transistors had the RT mobility in the range from 30000 to 36000 cm$^2$/Vs and operated in the near ballistic regime. It was established that 1/$f$ noise in such device is suppressed by a factor of ×5 − ×10 as compared to that in non-encapsulated graphene devices on Si/SiO$_2$. Considering that $h$-BN is chemically stable and produces strong positive effect on mobility in graphene channel, our finding that the $h$-BN capping and barrier layers result in significant reduction of 1/$f$ noise adds an extra merit to practical electronic applications of graphene-based heterostructures.

*Acknowledgements*

The work at UC Riverside was supported, in part, by the Semiconductor Research Corporation (SRC) and Defense Advanced Research Project Agency (DARPA) through STARnet Center for Function Accelerated nanoMaterial Engineering (FAME) and by the National Science Foundation (NSF) project Graphene Circuits for Analog, Mixed-Signal, and RF Applications (NSF CCF-1217382). The work at RPI was supported by the Army Research Office (Program Manager: Dr. Meredith Reed).

**FIGURE CAPTIONS**

**Figure 1:** (a) Optical microscopy image of an assembled *h*-BN - graphene - *h*-BN stack. (b) Raman spectrum of an exfoliated single layer graphene flake. The absence of the disorder D peak indicates the high quality of graphene.

**Figure 2:** (a) Schematics of *h*-BN-G-*h*-BN HFET. Note the structure the "one-dimensional" contact to the fully encapsulated graphene layer. (b) Optical microscopy image of a representative graphene encapsulated HFET.

**Figure 3:** Current – voltage transfer characteristics of *h*-BN-G-*h*-BN HFETs. The source-drain voltage is 10 mV.

**Figure 4:** Normalized noise spectrum density in *h*-BN-G-*h*-BN HFET as a function of frequency for several values of the back-gate bias $V_G$. Note that $V_G$=-7.75 corresponds to the Dirac point.

**Figure 5:** Noise amplitude as a function of the gate bias with respect to the Dirac point, $V_{GS}$-$V_D$ for *h*-BN-G-*h*-BN HFET. The results are shown for the device with the largest channel dimensions. The data for conventional non-encapsulated graphene FET on Si/SiO$_2$ wafer from Ref. [28] are also shown for comparison.

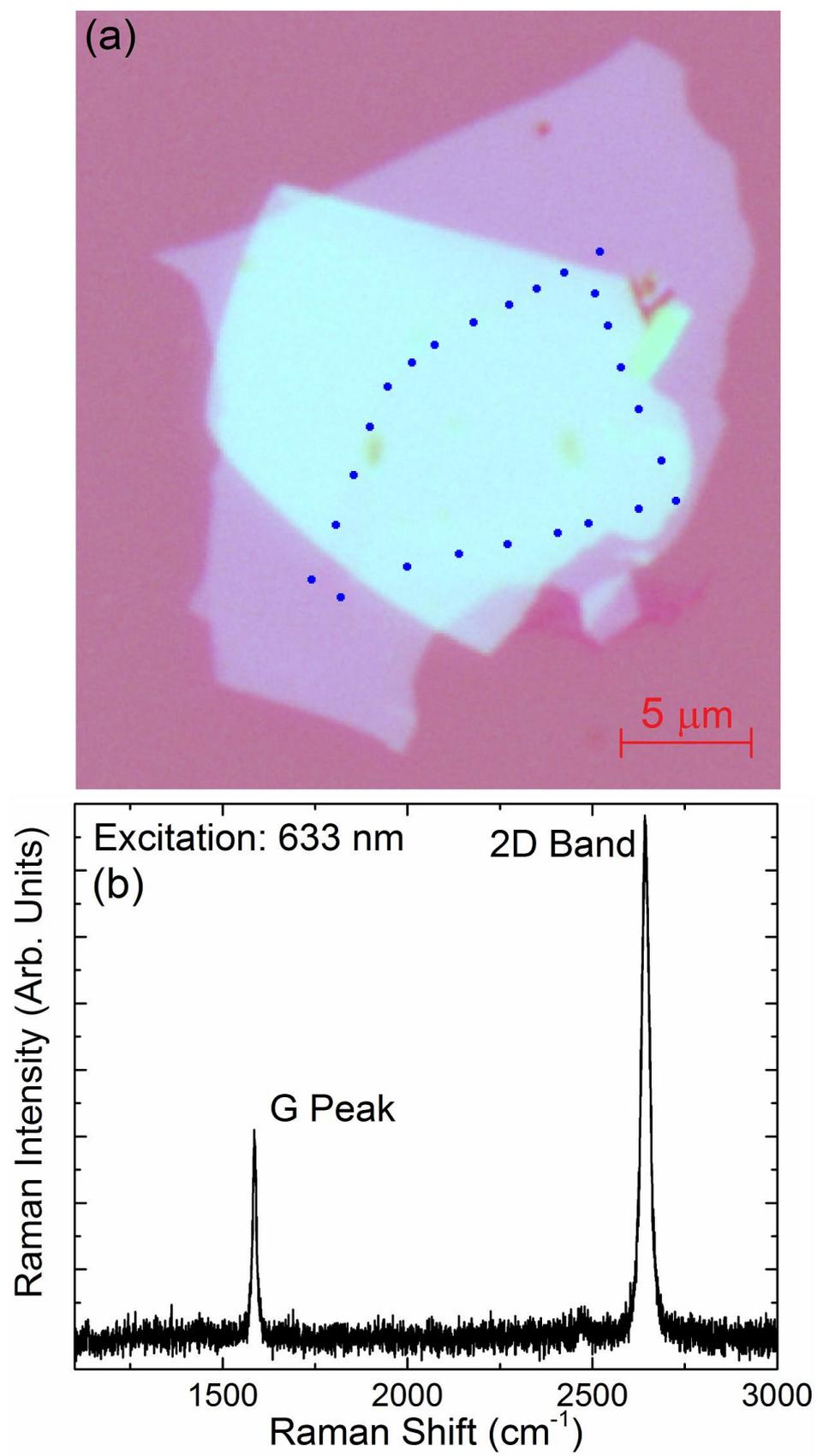

Figure 1

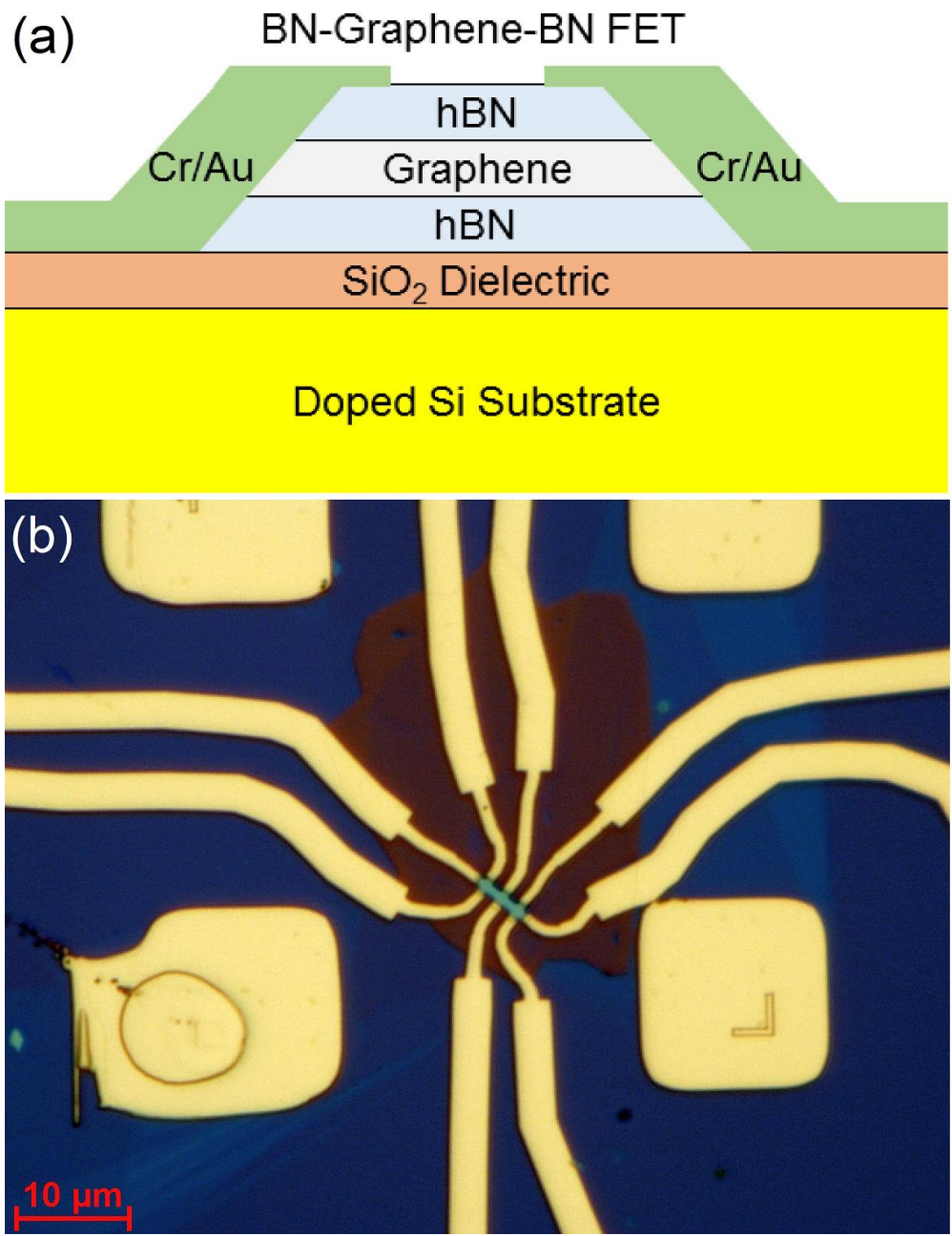

Figure 2

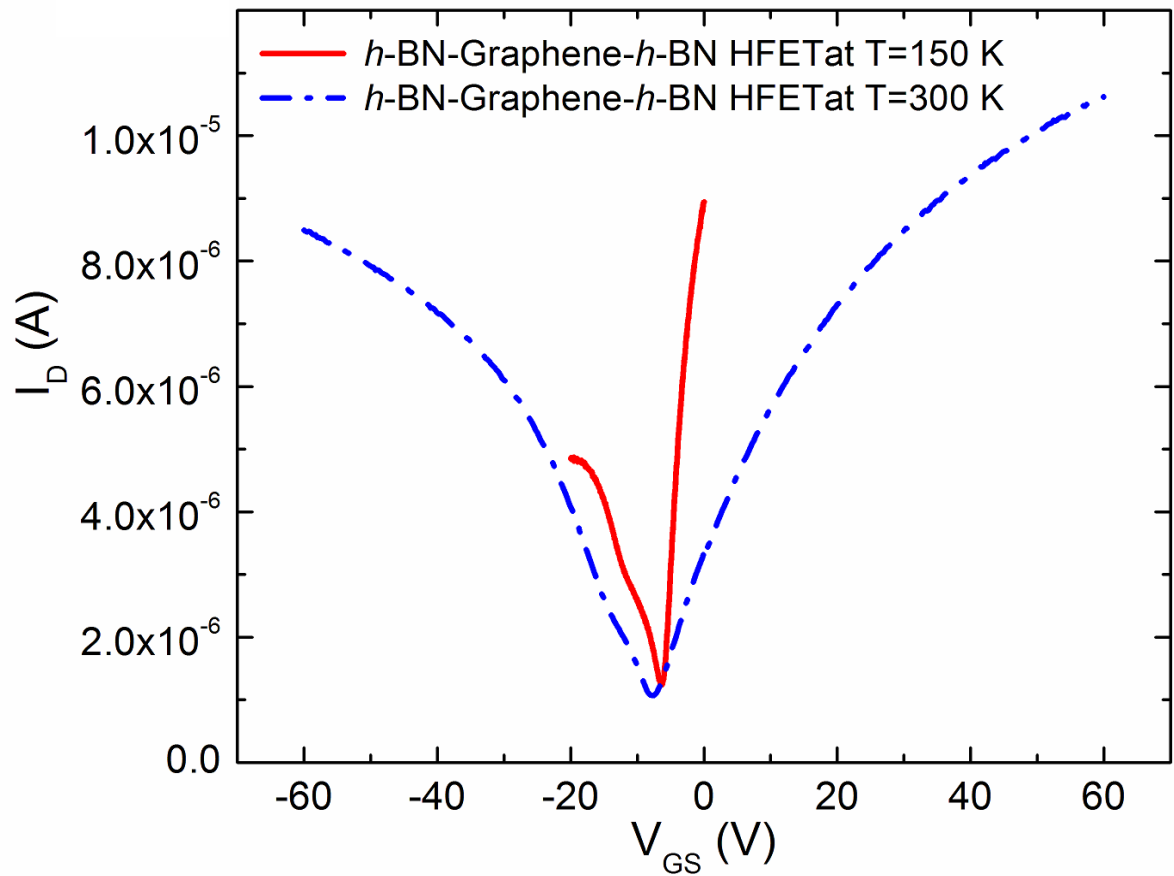

Figure 3

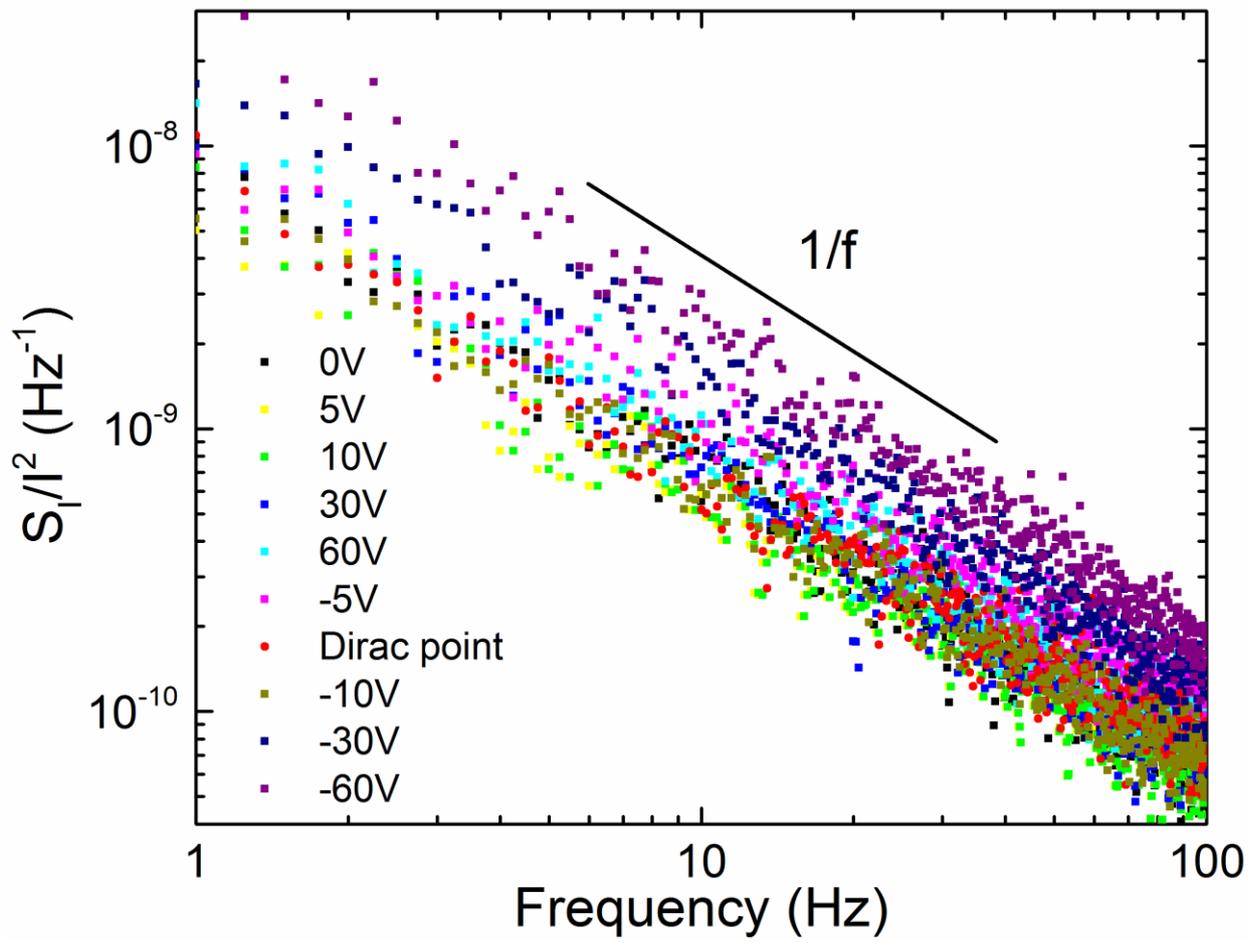

Figure 4

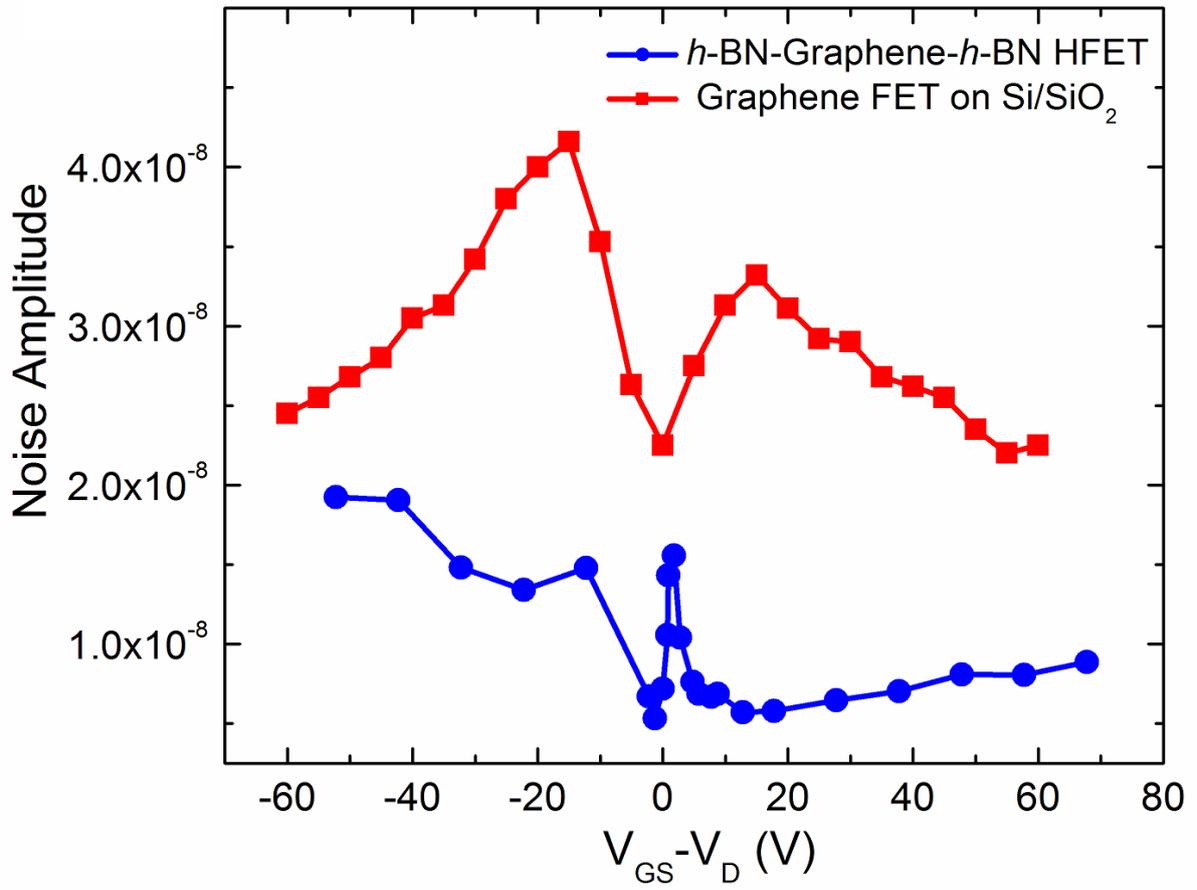

Figure 5